\documentclass[preprint,showpacs,preprintnumbers,amsmath,amssymb]{revtex4}
\usepackage{graphics,epsfig,subfigure}
\usepackage{epstopdf}
\usepackage{color}

\begin{document}
\renewcommand{\baselinestretch}{1.3}
\newcommand\be{\begin{equation}}
\newcommand\ee{\end{equation}}
\newcommand\ba{\begin{eqnarray}}
\newcommand\ea{\end{eqnarray}}
\newcommand\nn{\nonumber}
\newcommand\fc{\frac}
\newcommand\lt{\left}
\newcommand\rt{\right}
\newcommand\pt{\partial}

\title{Tensor perturbations of Palatini $f(\mathcal{R})$-branes}
\author{Bao-Min Gu\footnote{gubm09@lzu.edu.cn},
        Bin Guo\footnote{guob12@lzu.edu.cn},
        Hao Yu\footnote{yuh13@lzu.edu.cn}
        and Yu-Xiao Liu\footnote{liuyx@lzu.edu.cn, corresponding author}}

\affiliation{Institute of Theoretical Physics, Lanzhou University, Lanzhou 730000, China.}

\begin{abstract}
We investigate the thick brane model in Palatini $f(\mathcal{R})$ gravity. The brane is generated by a real scalar field with a scalar potential. We solve the system analytically and obtain a series of thick brane solutions for the $f(\mathcal{R})=\mathcal{R}+\alpha \mathcal{R}^2$-brane model. It is shown that tensor perturbations of the metric are stable for $df({\mathcal{R}})/d{\mathcal{R}}>0$.
For nonconstant curvature solutions, the graviton zero mode can be localized on the brane, which indicates that the four-dimensional gravity can be recovered on the brane. Mass spectrum of graviton KK modes and their corrections to the Newtonian potential are also discussed.
\end{abstract}


\maketitle



\section{Introduction}

Recent observations indicate that to explain the motions of galaxy and the cosmic speed-up, dark matter and dark energy should be included in the framework of general relativity, and they sum up 96\% of the total energy content. This inevitably puts forward a challenge to general relativity. General relativity actually has passed all tests at solar-system scales only, but not all length scales. So considering alternative theories of gravity is a reasonable choice.

In fact, Einstein considered a new approach to variation after he found general relativity, nowadays known as Palatini variation. Unlike the conventional metric variation, the Palatini variation assumes that both the metric and connection are independent variables, and thus abandons the priori of metric. Under Palatini variational principle, two field equations can be obtained by varying with respect to the metric and  connection. In Palatini theories, the matter actions are assumed to be independent of the connection. For the Einstein-Hilbert action, these two variations are equivalent~\cite{Wald1984a}.
However, for the action of a general $f(R)$ gravity, these two variations would lead to two very different theories: the metric and Palatini $f(R)$ theories.
It is well known that the metric
$f(R)$ theory leads to fourth-order field equations,
while the Palatini $f(\mathcal{R})$ theory leads to second-order ones.
In recent years the Palatini $f(\mathcal{R})$ theories of gravity have attracted
great interest since they are expected to have good descriptions of
the phenomenons of our universe~\cite{Olmo2011b, Olmo2005b, Fay2007a, Amarzguioui2006, Barragan2010a, Tsujikawa2008, Koivisto2007a, Sotiriou2006, Flanagan2004, Meng2004a, Meng2004,Allemandi:2005qs}, and they are different from the results of the metric $f(R)$ gravity~\cite{Nojiri2003,Hu2007}.  In Ref.~\cite{Vollick2003}, it was shown that $1/\mathcal{R}$ correction
to the Einstein-Hilbert action in Palatini formalism offers an alternative
explanation for late time acceleration. The models with the addition of
both positive and negative powers of scalar curvature were considered in Ref.~\cite{Sotiriou2006a}.
It was shown that these models of modified gravity may account for both early time inflation
and late time accelerated expansion.

On the other hand, it has been shown that brane world theories may address some open problems in particle physics and phenomenology such as the hierarchy problem and cosmological constant problem
~\cite{Akama1982,Arkani-Hamed1998a,Arkani-Hamed1999a,Antoniadis1998a,
Randall1999,Randall1999a,Maartens2004a}.
Inspired by these theories, thick brane world models~\cite{Csaki2000a,Gremm2000,Gremm2000a,Kobayashi:2001jd} have been considered in both general relativity and modified gravity theories. In these models, a lot of interesting  structures have been found~\cite{Dzhunushaliev2010a,Liu2011a,Fu2014a}. Another reason to consider thick brane models is that most of the thin brane models constructed in modified gravity with higher derivatives cannot be solved. As in the Randall-Sundrum model~\cite{Randall1999a}, one of the issues of thick brane model is the localization of the graviton zero mode, which is related to the recovery of the four-dimensional gravity. Besides, the stability problem of the system is also very important. Usually, the graviton massive KK modes are suppressed on the brane, but they do contribute to the Newtonian potential, and this provides an approach to detect the extra dimensions.

The structure of a brane world model is determined by the gravity model. As is well known, the metric $f(R)$ gravity theories modify the gravitational sector of the Einstein equations. For the brane world models constructed in the metric $f(R)$ gravity, see Refs.~\cite{Liu2011a,Zhong2011,Afonso2007a,Bazeia:2007jj,Dzhunushaliev2010,HoffdaSilva:2011si,Liu:2011am,Bazeia2013b,Bazeia2014a}. In contrast to the metric $f(R)$ gravity, the Palatini $f(\mathcal{R})$ gravity is equivalent to general relativity with a modified source.
In this paper, we expect that the thick brane model in Palatini $f(\mathcal{R})$ gravity has some interesting features, in particular the solutions and the tensor perturbations.

In Ref.~\cite{Bazeia2014b}, thick brane solutions in Palatini $f(\mathcal{R})$ gravity were obtained under first-order framework and perturbative approach. We try to get analytic solutions of the thick brane model for general $f(\mathcal{R})$ with constant curvature and exact solutions of the Palatini $f(\mathcal{R})=\mathcal{R} + \alpha\mathcal{R}^{2}$ gravity with nonconstant curvature in this paper. In section~\ref{secModel}, we first review the Palatini $f(\mathcal{R})$ gravity and set up our Palatini $f(R)$-brane model. Then we derive the second-order field equations in five dimensions for our model and show how to solve the field equations analytically. In section~\ref{SecFluctuations}, we study the gravitational fluctuations and stability problems. The localization of the graviton zero mode and the corrections to Newtonian potential of massive KK modes are also discussed. The discussion and conclusions are given in section~\ref{secConclusion}.

\section{The model}
\label{secModel}

 In this section, we consider the general $f(\mathcal{R})$ model in five-dimensional spacetime in Palatini formalism. The action takes the form
 \be
 S_{\texttt{Pal}}=\frac{1}{2\kappa_{5}^{2}} \int d^5 x \sqrt{-g}f(\mathcal{R}(g,\Gamma))+S_{\texttt{M}}(g_{M N},\Psi),\label{action of Pal. f(R)}
 \ee
 where $\kappa_{5}^{2}=1/M^{3}_{*}$ with $M_{*}$ the fundamental scale, $g$ is the determinant of the metric, and $S_{\texttt{M}}(g_{M N},\Psi)= \int d^5 x \sqrt{-g} L_{\texttt{M}}(g_{M N},\Psi)$ is the action for ordinary matter that only couples to the metric. In this paper, capital Latin letters $M,N,\cdots$ denote the five-dimensional coordinate indices $0,1,2,3,5$ and Greek letters $\mu,\nu,\cdots$ denote the four-dimensional coordinate indices $0,1,2,3$. $\mathcal{R}(g,\Gamma)=g^{M N}\mathcal{R}_{M N}$ is the Ricci scalar constructed by the independent connection $\Gamma$. In Palatini $f(\mathcal{R})$ gravity theories, the main feature is that both the metric and the connection are assumed to be independent variables. It is very different from general relativity and other metric theories. We will see that this set-up leads to special physics.
 Varying with respect to the metric and the connection, respectively, one gets the following two field equations:
\ba
 f_{\mathcal{R}}\mathcal{R}_{M N}-\frac{1}{2}f g_{M N}
     &=&  \kappa_{5}^{2}T_{M N},
     \label{field equation of metric}\\
 \tilde{\nabla}_{A}\left(\sqrt{-g}f_{\mathcal{R}}g^{M N}\right)
     &=&0,     \label{field equation of connection}
\ea
 where $f_{\mathcal{R}}\equiv{df}/{d\mathcal{R}}$,
 $T_{M  N}$ is the energy-momentum tensor,
 and $\tilde\nabla$ is the covariant derivative defined with the connection $\Gamma$.
 Note that $\tilde\nabla$ is not compatible with the metric, which implies that $\tilde{\nabla}_{A}g_{MN}\neq0$ unless $f(\mathcal{R})=\mathcal{R}$. Actually, Eq. (\ref{field equation of connection}) defines the auxiliary metric in Palatini $f(R)$ gravity. If we define
\be
  \sqrt{-q}q^{M N}\equiv \sqrt{-g}f_{\mathcal{R}}g^{M N},
  \label{definition of auxiliary metric}
\ee
 then we have $\tilde\nabla_{A}(\sqrt{-q}q^{M N})=0$.
 It is similar to the equation $\nabla_{A}(\sqrt{-g}g^{M N})=0$ (it is also equivalent to $\nabla_{A}g^{MN}=0$) in general relativity. At this point, we can say $\tilde\nabla$ is compatible with the auxiliary metric $q^{M N}$. According to this definition, we obtain
\be
 q^{M N}=f_{\mathcal{R}}^{-{2}/{3}}g^{M N},\quad
 q_{M N}=f_{\mathcal{R}}^{{2}/{3}}g_{M N}.
              \label{definition of metric q}
\ee
 Obviously, $q_{M N}$ is just the conformally transformed metric. With this metric, one can express the independent connection as
\ba
 \Gamma_{M N}^{A}&=&\frac{1}{2}q^{A B}(\partial_{M}q_{N B}+\partial_{N}q_{M B}-\partial_{B}q_{M N})\nn\\&=&\left\{^{A}_{M N}\right\}+C^{A}_{M N},
 \label{definition of independent connection}
\ea
 where $\left\{^{A}_{M N}\right\}$ is the Christoffel symbol and $C^{A}_{M N}$ is a well-defined tensor. When $f(\mathcal{R})=\mathcal{R}$, we have $C^{A}_{M N}=0$, and the theory would reproduce general relativity.

 The expression (\ref{definition of independent connection}) indicates that we are able to eliminate the independent connection $\Gamma$ from the field equations. If this is done, then we would get one field equation which only relies on metric dynamically. With the following relations
 \ba
 \mathcal{R}_{M N}&=&R_{M N}(g)-\frac{1}{3f_{\mathcal{R}}}
 \left(3\nabla_{M}\nabla_{N}f_{\mathcal{R}}+g_{M N}\nabla_{A}\nabla^{A}f_{\mathcal{R}}\right)
 +\frac{4}{3f_{\mathcal{R}}^{2}}\nabla_{M}f_{\mathcal{R}}\nabla_{N}f_{\mathcal{R}},
 \label{expression of transformed Ricci tensor}\\
 \mathcal{R}&=&R-\frac{8}{3f_{\mathcal{R}}}\nabla_{A}\nabla^{A}f_{\mathcal{R}}
 +\frac{4}{3f_{\mathcal{R}}^{2}}\nabla_{A}f_{\mathcal{R}}\nabla^{A}f_{\mathcal{R}},
 \label{expression of transformed Ricci scalar}
 \ea
 where  ${R}_{M N}(g)$ and ${R}$ are the Ricci tensor and Ricci scalar constructed by the spacetime metric $g_{MN}$, respectively,
 we can transform the Eq. (\ref{field equation of metric}) into the following one
\ba
 G_{M N}&=&\frac{\kappa_{5}^{2} T_{M N}}{f_{\mathcal{R}}}-
 \frac{1}{2}g_{M N}\left(\mathcal{R}-\frac{f}{f_{\mathcal{R}}}\right)+
 \frac{1}{f_{\mathcal{R}}}\left(\nabla_{M}\nabla_{N}-g_{M N}\nabla_{A}\nabla^{A}\right)
 f_{\mathcal{R}}\nn\\
 &&-\frac{4}{3f_{\mathcal{R}}^{2}}
 \left(\nabla_{M}f_{\mathcal{R}}\nabla_{N}f_{\mathcal{R}}
 -\frac{1}{2}g_{M N}\nabla_{A}f_{\mathcal{R}}\nabla^{A}f_{\mathcal{R}}\right),
 \label{modified Einstein equation}
\ea
 where $G_{M N}=R_{MN}-\frac{1}{2}R g_{MN}$ is the Einstein tensor. Furthermore, from Eq. (\ref{field equation of metric}), we have
\be
f_{\mathcal{R}}\mathcal{R}-\frac{5}{2}f=\kappa_{5}^{2} T\label{trace of Einstein eq.},
\ee
 which shows that $\mathcal{R}$ is related to the trace of energy-momentum tensor  algebraically. Thus, all of the quantities such as $\mathcal{R}$, $f(\mathcal{R})$, and $f_{\mathcal{R}}$ can be expressed by $T$. At this point, we have successfully eliminated  the auxiliary metric $q^{M N}$ or the connection $\Gamma$ from the field equations, and the dynamical variable of the field equations (\ref{modified Einstein equation}) is the spacetime metric $g_{MN}$. 
 The implication of Eq. (\ref{modified Einstein equation}) is clear so far: it is the Einstein equation with a modified source, and the effective energy-momentum tensor is defined by the right hand side of Eq. (\ref{modified Einstein equation}).
 It can be seen that the Palatini $f(\mathcal{R})$ gravity is equivalent to a metric theory with a modified source. For more details about the $f(R)$ gravity, see Refs.~\cite{Sotiriou2010,Capozziello2011a,DeFelice2010,Nojiri2011}.

 In this paper we consider the $f(\mathcal{R})$ brane model with a scalar field presented in the five-dimensional background spacetime. The background metric with four-dimensional Poincar$\acute{e}$ symmetry is assumed as
\be
d s^2=a^2(y)\eta_{\mu\nu}d x^{\mu}d x^{\nu}+d y^{2},\label{background metric}
\ee
 where $a(y)$ is the warp factor. The Lagrangian of the scalar field is assumed as $\mathcal{L}_{\phi}=-\frac{1}{2}\partial_{M}\phi\partial^{M}\phi-V(\phi)$. The corresponding energy-momentum tensor and equation of motion of the scalar field are
\ba
 T_{M N}&=&\partial_{M}\phi\partial_{N}\phi-g_{M N}\left(\frac{1}{2}\partial_{A}\phi\partial^{A}\phi+V(\phi)\right),
 \label{energy-momentum tensor}\\
 \Box^{(5)}{\phi}&=& V_{\phi}.\label{EOM of scalar}
\ea
 For static brane solution, the scalar field is a function of $y$, namely $\phi=\phi(y)$. Then, with the metric (\ref{background metric}), the explicit forms of the above two equations are given by
\ba
 T_{\mu\nu}&=&-a^2 \Big(\frac{1}{2}\phi'^2+V\Big) \eta_{\mu\nu},
 \label{energy-momentum tensor1}\\
 T_{55}&=&\frac{1}{2}\phi'^2-V,
 \label{energy-momentum tensor2}\\
 V{'}&=&\phi{''}\phi{'}+4\frac{a{'}}{a}\phi'^{2},\label{EOM of scalar1}
\ea
 where the prime represents the derivative with respect to the extra dimension coordinate $y$.
 To construct a thick brane world model, we expect to solve the system (\ref{modified Einstein equation}) and (\ref{EOM of scalar1}).
Usually, we can obtain topologically nontrivial solutions by introducing a superpotential~\cite{Gremm2000, Fu2011a, Chen2013} or giving a scalar potential such as the $\phi^{4}$ or other models.
 However, in our case this does not work because of the complex expression of the right hand side of (\ref{modified Einstein equation}). To solve this system, we consider Eqs. (\ref{field equation of metric}) and (\ref{definition of auxiliary metric}) instead of Eq. (\ref{modified Einstein equation}). With the relation (\ref{definition of auxiliary metric}) between $q_{M N}$ and $g_{M N}$, it is convenient to assume the auxiliary metric as
\be
 d\tilde{s}^2=u^2(y)\eta_{\mu\nu}dX^{\mu}dX^{\nu}+\frac{u^{2}(y)}{a^{2}(y)}dY^{2}.
\ee
 Then Eqs. (\ref{field equation of metric}) and (\ref{definition of auxiliary metric}) are reduced to
\ba
 \left(6\frac{u'^{2}}{u^{2}}-3\frac{a'}{a}\frac{u'}{u}-3\frac{u''}{u}\right)
 f_{\mathcal{R}}&=&\kappa_{5}^{2}\phi'^{2},
 \label{compent equation a}\\
 5f_{\mathcal{R}}\left(\frac{a'}{a}\frac{u'}{u}+\frac{u''}{u}\right)
 -2f_{\mathcal{R}}\frac{u'^{2}}{u^{2}}+f(\mathcal{R})&=&2\kappa_{5}^{2} V,
 \label{compent equation b}
\ea
 and
\be
 f_{\mathcal{R}}=\left(\frac{u}{a}\right)^3,
 \label{definition equation of q}
\ee
 respectively.

\subsection{Constant curvature solutions}

 Now we have four equations (\ref{EOM of scalar}), (\ref{compent equation a}), (\ref{compent equation b}), and (\ref{definition equation of q}). However, Eqs. (\ref{EOM of scalar}), (\ref{compent equation a}), and (\ref{compent equation b}) are not independent because of the conservation of $T_{M N}$. To solve the system we need a constraint. Obviously, different constraints lead to different results. We first consider the case in which $\mathcal{R}(\Gamma)$ is a constant. According to Eq. (\ref{definition of metric q}), it is straightforward to conclude that $R(g)$ is also constant. Thus, the solutions are the same as in metric $f(R)$ gravity with constant $R(g)$~\cite{Zhong2011}. The solutions are listed as follows.
\begin{itemize}
\item For $AdS_{5}$, $\mathcal{R}(\Gamma)=R(g)=-20{\gamma}^{2}(\gamma>0)$ and $f_{\mathcal{R}}<0$, we have
\ba
 a(y)&=&\text{cosh}^{{2}/{5}}\left(\frac{5{\gamma}y}{2}\right),\nn\\
 \phi(y)&=&\pm2\sqrt{\frac{6|f_{\mathcal{R}}|}{5\kappa_{5}^{2}}}
 \text{arctan}\left(\text{tanh}\left(\frac{5{\gamma}y}{4}\right)\right),\label{AdSsolution}\\
 V(y)&=&V_{0}+\frac{9{\gamma}^{2}|f_{\mathcal{R}}|}{4\kappa_{5}^{2}}
 \text{sin}^{2}\left(\sqrt{\frac{5\kappa_{5}^{2}}{6|f_{\mathcal{R}}|}}\phi\right),\nn
\ea
where $V_{0}=({2f-25{\gamma}^{2}|f_{\mathcal{R}}|})/{4\kappa_{5}^{2}}$.

\item For $dS_{5}$, $\mathcal{R}(\Gamma)=R(g)=20{\gamma}^{2}$ and $f_{\mathcal{R}}>0$,
\ba
 a(y)&=&\text{cos}^{{2}/{5}}\left(\frac{5{\gamma}y}{2}\right),\nn\\
 \phi(y)&=&\pm\sqrt{\frac{6f_{\mathcal{R}}}{5\kappa_{5}^{2}}}
 \text{arctanh}\left(\text{sin}\left(\frac{5{\gamma}y}{2}\right)\right), \label{dSsolution}\\
 V(y)&=&V_{0}-\frac{9{\gamma}^{2}f_{\mathcal{R}}}{4\kappa_{5}^{2}}
 \text{sinh}^{2}\Bigg(\sqrt{\frac{5\kappa_{5}^{2}}{6f_{\mathcal{R}}}}\phi\Bigg),\nn
\ea
 where $V_{0}=({2f-25{\gamma}^{2}f_{\mathcal{R}}})/{4\kappa_{5}^{2}}$.
\end{itemize}

 The $AdS_{5}$ solution supports a warp factor which diverges at infinity. However, it
 can be checked that for an observer located at $y=0$, photons coming from infinity cost
 finite time to reach $y=0$. Therefore, there are no event horizons here. For the $dS_{5}$ solution, the extra dimension should be restricted to the interval $-{\pi}/{5{\gamma}}< y < {\pi}/{5{\gamma}}$ and there are also no event horizons.
 Some more discussions are given in section~\ref{Stability}.

\subsection{Nonconstant curvature solutions}

 For nonconstant $\mathcal{R}(\Gamma)$, the system becomes complex. For the metric $f(R)$ gravity, it is the metric that involves higher
 derivatives,  so the brane solutions can be obtained by assuming the solution of the warp factor~\cite{Liu2011a}.
 However, this is not a good choice for our case.

Here, we consider the $f(\mathcal{R})=\mathcal{R}+\alpha \mathcal{R}^2$ model, for which Eq. (\ref{definition equation of q}) becomes
\be
 1-8\alpha\left(2\frac{a'}{a}\frac{u'}{u}
 +2\frac{u''}{u}+\frac{u'^{2}}{u^2}\right)
 =\left(\frac{u}{a}\right)^3.
 \label{definition equation of q_2}
\ee
Note that the system can be greatly simplified if we impose a good relation between $u(y)$ and $a(y)$. For this purpose, we assume $u(y)=c_{1}a^{n}(y)$ with $n\ne 0$. Then Eq.
(\ref{definition equation of q_2}) reduces to
\be
 1-24n^{2}\alpha\frac{a'^2}{a^2}-16n\alpha\frac{a''}{a}
 -c_{1}^{3}a^{3(n-1)}=0. \label{EqofModel}
\ee
 When $n=1$, we get the solutions (\ref{AdSsolution}) and (\ref{dSsolution}), which are  constant curvature solutions. So we consider the case $n\neq1$, for which Eq.~(\ref{EqofModel}) supports the following solution of the warp factor:
\be
 a(y)=\text{sech}^{\frac{2}{3(n-1)}}(ky) \label{ay}
\ee
 with $k=\frac{3(n-1)}{\sqrt{32n(3n+2)\alpha}}$, and $c_1$ in (\ref{EqofModel}) is fixed as $c_1=(\frac{6n-1}{3n+2})^{1/3}$. Note that this solution is valid for $\alpha\neq 0$ ($\alpha=0$ corresponds to general relativity and can be solved with superpotential method). Thus the potential $V(y)$, scalar field $\phi(y)$, and energy density $\rho(y)$ are given by
\ba
 V(y)&=&\frac{(3n+1)(6n-1)}{32(3n+2)^2 \kappa_{5}^{2}\alpha}
 \text{sech}^4(ky) + \frac{(3n+5)(6n-1)}{16(3n+2)^2 \kappa_{5}^{2}\alpha}
 \text{sech}^2(ky) +\Lambda_5,
 \label{sol of potential}\\
 \phi(y)&=&\sqrt{\frac{2n(6n-1)}{3(3n+2)(n-1)\kappa_{5}^2}}
  \Bigg[i \sqrt{3} ~\text{E}\left(ik y, \frac{2}{3}\right)
  -i \sqrt{3} ~\text{F}\left(ik y, \frac{2}{3}\right)\nonumber\\
   &&+ \sqrt{2+\text{cosh}(2k y)}\text{tanh}(k y)\Bigg],
 \label{sol of scalar}\\
 \rho(y)&=&\frac{(3n-1)(6n-1)}{16(3n+2)^2 \kappa_{5}^{2}\alpha}
 \text{sech}^4(ky) + \frac{(3n+1)(6n-1)}{8(3n+2)^2 \kappa_{5}^{2}\alpha}
 \text{sech}^2(ky),
 \label{sol of energy density}
\ea
 where $\Lambda_5 =-{1}/{8\kappa_{5}^{2}\alpha}$, $\text{F}(y,m)$ and $\text{E}(y,m)$ are the incomplete elliptic integrals of the first and second kinds, respectively.

 The scalar field involves Elliptic integrals, and we fail to give the expression of the potential $V(\phi)$. It can be seen that our solutions are determined by two parameters $n$ and $\alpha$. In order to get an asymptotic $AdS_5$ solution, we require $\alpha>0$. The solution of the scalar field indicates that $n$ should be restricted to the interval
 $n<-{2}/{3}$ or $0<n<{1}/{6}$ or $n>1$. 
Besides, any solution with $0<n<{1}/{6}$ leads to negative energy density, so we exclude these solutions. On the other hand, for an observer located at the origin of the extra dimension, photons coming from infinity cost
finite time to reach the origin for the warp factor with $n<-2/3$.
We will see in section~\ref{subsecMassiveKKmodes} that the solutions with $n<-2/3$ have some interesting features.
\begin{figure*}[htb]
\begin{center}
\includegraphics[width=6.5cm,height=5cm]{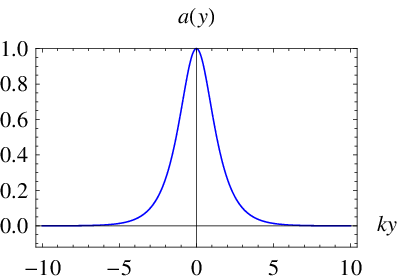}
\includegraphics[width=7cm,height=5cm]{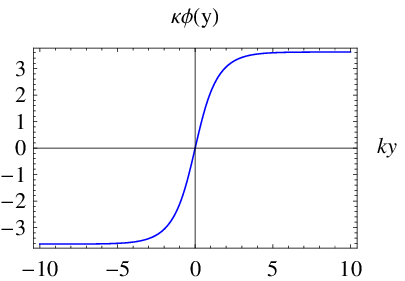}
\includegraphics[width=7cm,height=5cm]{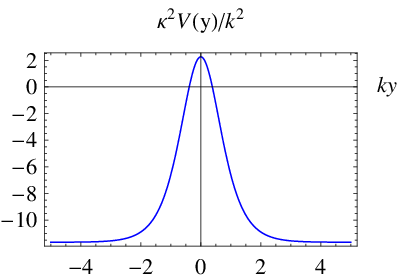}
\includegraphics[width=7cm,height=5cm]{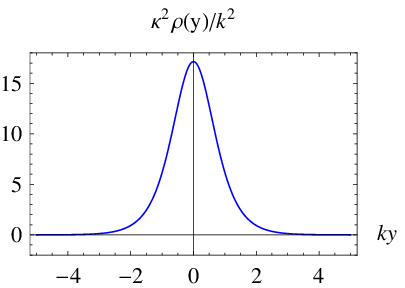}
\end{center}
\caption{The shapes of the $a(y)$, $\kappa_5\phi(y)$, $\kappa_{5}^2 V(y)/k^2$, and $\kappa_{5}^2\rho(y)/k^2$ with respect to $ky$ for $n={5}/{3}$.}\label{fig1}
\end{figure*}
Here, we note that the brane world solution for $n={5}/{3}$ becomes simpler,
and it reads
\ba
 a(y)&=& \text{sech}(ky),
 \label{sol of warp factor}\\
 \phi(y)&=&\sqrt{\frac{15}{7\kappa_{5}^2}}
  \Bigg[i \sqrt{3}~ \text{E}\left(ik y, \frac{2}{3}\right)
  -i \sqrt{3} ~\text{F}\left(ik y, \frac{2}{3}\right)
  \nonumber\\ &&+ \sqrt{2+\text{cosh}(2ky)}\,\text{tanh}(ky)\Bigg],\\
 V(y)&=&\frac{5k^2}{\kappa_{5}^{2}} \left( \frac{9}{14} \text{sech}^4(ky)
  + \frac{15}{7}\text{sech}^2(ky)
  - \frac{7}{3}\right),
\ea
 where $k=\sqrt{\frac{3}{280\alpha}}$.
 It is clear that $a(\pm\infty)\rightarrow0$, thus $|\phi(\infty)|\rightarrow$constant, and $V(\pm\infty)\rightarrow \Lambda_5$. We show the plots of $a(y)$, $\phi(y)$, and $V(y)$ in figure~\ref{fig1}.

 The thickness of the brane, $1/k$, is determined by the parameter $\alpha$. Another feature of our solution is that the cosmological constant $\Lambda_5=-{1}/{8\kappa_{5}^{2}\alpha}$ is independent of $n$. Equation (\ref{sol of energy density}) implies that the energy density peaks at $y=0$, and it does not dissipate with time. Furthermore, the Ricci scalar $R(g)$ is given by
\be
    R(g)=  -\frac{8 k^2 [1-6 n+5 \cosh(2 k y)] \text{sech}^2(k y)}
              {9 (n-1)^2}.
\ee
 As $y\rightarrow\pm\infty$,  $R(g)\rightarrow-80 k^2/9 (n-1)^2<0$. This is consistent with the fact that the spacetime far away from the brane is asymptotically $AdS_{5}$.

\section{Gravitational fluctuations}
\label{SecFluctuations}

The fluctuations $\delta g_{MN}$ of the background metric (\ref{background metric})  can be decomposed as the transverse-traceless (TT) tensor mode, transverse vector modes, and scalar modes.
It can be shown that the transverse vector modes and scalar modes are decoupled with the TT tensor modes.
In this paper we would like to investigate stability and localization of the TT tensor fluctuations of the background metric (\ref{background metric}), whose KK modes are related to the four-dimensional gravitons and Newtonian potential.

\subsection{Stability under tensor perturbations}\label{Stability}
In our case, the perturbed metric is
\be
 d s^2=a^2(y)(\eta_{\mu\nu}+h_{\mu\nu})d x^{\mu}d x^{\nu}+d y^{2},
\label{perturbed metric}
\ee
where the tensor fluctuations $h_{\mu\nu}$ satisfy the TT condition
\be
 \partial_{\mu}h^{\mu}_{~\nu}=0,\quad h\equiv\eta^{\mu\nu} h_{\mu\nu}=0. \label{TT-Condition}
\ee
 Thus we have
\be
\delta g_{\mu\nu}=a^2(y)h_{\mu\nu},\quad \delta g_{55}=\delta g_{\mu 5}=0.
\ee
 With the perturbed metric (\ref{perturbed metric}), to linear order, we get the perturbations of Ricci tensor and Ricci scalar
\ba
 \delta R_{\mu\nu}=&&\!\!\!\!\!\!\!\!\!\frac{1}{2}
 \left(\partial_{\sigma}\partial_{\nu}h^{\sigma}_{~\mu}
 +\partial_{\sigma}\partial_{\mu}h^{\sigma}_{~\nu}-\Box ^{(4)}h_{\mu\nu}
 -\partial_{\mu}\partial_{\nu}h\right)\nn \\&-&3a'^{ 2}h_{\mu\nu}-aa{''}h_{\mu\nu}-2aa{'}h_{\mu\nu}'-
 \frac{1}{2}a^{2}h_{\mu\nu}''-\frac{aa{'}h{'}}{2}\eta^{\mu\nu}, \label{perturbation of Ricci tensor}\\
 \delta R=&&\!\!\!\!\!\!\!\!\!\frac{1}{a^{2}}\left(\partial_{\mu}\partial_{\nu}h^{\mu\nu}
 -\Box^{(4)}h\right)-h{''}-\frac{5a{'}}{a}h{'},
 \label{perturbation of Ricci scalar}
\ea
 where $\Box^{(4)}=\eta^{\mu\nu}\partial_{\mu}\partial_{\nu}$ denotes the four-dimensional d'Alembert operator. Considering the TT condition (\ref{TT-Condition}), the perturbation of Ricci scalar vanishes, and
\be
 \delta R_{\mu\nu}=-\frac{1}{2}\Box^{(4)}h_{\mu\nu}-3a'^{ 2}h_{\mu\nu}-aa{''}h_{\mu\nu}-2aa{'}h_{\mu\nu}'-\frac{1}{2}a^{2}h_{\mu\nu}''.
\ee
 We immediately obtain the perturbed $\mu\nu$-components of Einstein tensor
\ba
 \delta G_{\mu\nu}=&&\!\!\!\delta\left(R_{\mu\nu}-\frac{1}{2}R g_{\mu\nu}\right)\nn\\=&&\!\!\!-\frac{1}{2}\Box^{(4)}h_{\mu\nu}
 +3a'^{2}h_{\mu\nu}+3aa{''}h_{\mu\nu}
 -2aa{'}h_{\mu\nu}'-\frac{1}{2}a^{2}h_{\mu\nu}''.
 \label{perturbation of Einstein tensor}
\ea

 On the other hand, the perturbations of the $\mu\nu$-components of the right hand side of field equation~(\ref{modified Einstein equation}) reads
\ba
 \delta G_{\mu\nu}=\Bigg[&-&\frac{\kappa^2}{f_{\mathcal{R}}}\left(\frac{1}{2}\phi'^{ 2}+V(\phi)\right)-\frac{1}{2}\left(\frac{\kappa_{5}^{2} T}{f_{\mathcal{R}}}+\frac{3f}{2f_{\mathcal{R}}}\right)\nn\\
 &-&4\frac{a{'}}{a }\frac{\partial_{y}f_{\mathcal{R}}}{f_{\mathcal{R}}}
 -\frac{\partial^{2}_{y}f_{\mathcal{R}}}{f_{\mathcal{R}}}
 +\frac{2}{3}\left(\frac{\partial_{y}f_{\mathcal{R}}}{f_{\mathcal{R}}}\right)^{2}\Bigg]h_{\mu\nu}
 +\frac{\partial_{y}f_{\mathcal{R}}}{2f_{\mathcal{R}}}h_{\mu\nu}'.
 \label{perturbation of effctive SE tensor}
\ea
 With Eqs. (\ref{perturbation of Einstein tensor}) and (\ref{perturbation of effctive SE tensor}) we get the following perturbed equation
\ba
 &-&\frac{1}{2}\Box^{(4)}h_{\mu\nu}+3a'^{ 2}h_{\mu\nu}+3aa{''}h_{\mu\nu}-2aa{'}h_{\mu\nu}'-\frac{1}{2}a^{2}h_{\mu\nu}''\nn\\
 =\Bigg[&-&\frac{\kappa_{5}^{2}}{f_{\mathcal{R}}}\left(\frac{1}{2}\phi'^{ 2}+V(\phi)\right)-\frac{1}{2}\left(\frac{\kappa_{5}^{2} T}{f_{\mathcal{R}}}+\frac{3f}{2f_{\mathcal{R}}}\right)\nonumber\\
 &-&4\frac{a{'}}{a }\frac{\partial_{y}f_{\mathcal{R}}}{f_{\mathcal{R}}}
 -\frac{\partial^{2}_{y}f_{\mathcal{R}}}{f_{\mathcal{R}}}
 +\frac{2}{3}\left(\frac{\partial_{y}f_{\mathcal{R}}}{f_{\mathcal{R}}}\right)^{2}\Bigg]h_{\mu\nu}
 +\frac{\partial_{y}f_{\mathcal{R}}}{2f_{\mathcal{R}}}h_{\mu\nu}'.
 \label{perturbed modified field equation}
\ea
Next, we simplify the above equation. From Eq. (\ref{modified Einstein equation}), we have
\ba
 \frac{g^{\alpha\beta}G_{\alpha\beta}}{4}=
 &-&\frac{\kappa_{5}^2}{f_{\mathcal{R}}}\left(\frac{1}{2}\phi'^{2}+V(\phi)\right)
 -\frac{1}{2}\left(\frac{\kappa_{5}^2 T}{f_{\mathcal{R}}}
 +\frac{3f}{2f_{\mathcal{R}}}\right)\nn\\
 &-&\frac{4a{'}\partial_{y}f_{\mathcal{R}}}{a f_{\mathcal{R}}}
 -\frac{\partial^{2}_{y}f_{\mathcal{R}}}{f_{\mathcal{R}}}
 +\frac{2}{3}\left(\frac{\partial_{y}f_{\mathcal{R}}}{f_{\mathcal{R}}}\right)^{2}
 +\frac{a{'}\partial_{y}f_{\mathcal{R}}}{a f_{\mathcal{R}}}.
 \label{counterterm}
\ea
 Now, it is straightforward to get the following simplified perturbed equation by substituting Eq. (\ref{counterterm}) into Eq. (\ref{perturbed modified field equation}):
\be
 \frac{1}{2}\Box^{(4)}h_{\mu\nu}
 +2aa{'}h_{\mu\nu}'
 -\frac{1}{2}a^{2}h_{\mu\nu}''
 =-\frac{\partial_{y}f_{\mathcal{R}}}{2f_{\mathcal{R}}}h_{\mu\nu}'
 +\frac{a{'}\partial_{y}f_{\mathcal{R}}}{a f_{\mathcal{R}}}h_{\mu\nu},
 \label{perturbed eq.1}
\ee
 where we have used the result $g^{\alpha\beta}G_{\alpha\beta}
 =12(a'^{2}+aa{''})$. The above perturbed equation is our main equation. For convenience, we will transform it to a  Schr\"odinger-like equation.
 However, the third term on the left hand side contains a factor $a^2$, which destroys the formalism of the Schr\"odinger-like equation. To eliminate it, we introduce a coordinate transformation
\be
d y=a d z.
\ee
 Under this transformation, the background metric~(\ref{background metric}) turns to be a conformally flat one. As a consequence,
\be
 \partial_{y}=\frac{\partial_{z}}{a},\quad
 a{'}=\partial_{y}a=\frac{\partial_{z}a}{a}.
\ee
 Then Eq.~(\ref{perturbed eq.1}) becomes
\be
 \left[\Box^{(4)}-\partial^{2}_{z}+\left(3\frac{\partial_{z}a}{a}
 +\frac{\partial_{z}f_{\mathcal{R}}}{f_{\mathcal{R}}}\right)\partial_{z}\right]h_{\mu\nu}=0.
 \label{perturbed eq.2}
\ee
 By defining $\tilde{h}_{\mu\nu}=B(z)h_{\mu\nu}$ with $B(z)=a^{{3}/{2}}f_{\mathcal{R}}^{{1}/{2}}$, the equation of $\tilde{h}_{\mu\nu}$ reads
\be
 \left(\Box^{(4)} - \partial^{2}_{z} + \frac{\partial^{2}_{z}B}{B}\right)\tilde{h}_{\mu\nu}=0\label{perturbed eq.3}.
\ee
Now we introduce the KK decomposition $\tilde{h}_{\mu\nu}(x^{\sigma},z)=\varepsilon_{\mu\nu}(x^\sigma)\Psi(z)$. Then we get two equations:
\ba
 \left(\Box^{(4)}+m^2\right)\varepsilon_{\mu\nu}(x^\sigma)&=&0,\\
 \left(- \partial^{2}_{z} + \frac{\partial^{2}_{z}B}{B}\right)\Psi(z)&=&m^2\Psi(z). \label{SchrodingerEq}
\ea
It is clear that the equation for the function $\Psi(z)$ is a Schr\"odinger-like equation with the effective potential given by
\be
 \mathcal{W}(z)=\frac{\partial^{2}_{z}B}{B}=\frac{3\partial_{z}^{2}a}{2a}
    + \frac{\partial_{z}^{2}f_{\mathcal{R}}}{2f_{\mathcal{R}}}
    + \frac{3}{4}\left(\frac{\partial_{z}a}{a}\right)^{2}
    -\left(\frac{\partial_{z}f_{\mathcal{R}}}{2f_{\mathcal{R}}}\right)^{2}
    +\frac{3\partial_{z}a}{2a}\frac{\partial_{z}f_{\mathcal{R}}}{f_{\mathcal{R}}}.
 \label{Wz}
\ee
It is easy to show that Eq. (\ref{SchrodingerEq}) can be factorized as
\be
 \left(\partial_{z} + \frac{\mathcal{A}}{2}\right)\left(-\partial_{z} + \frac{\mathcal{A}}{2}\right)\Psi(z)=m^2 \Psi(z),
\ee
where $\mathcal{A}=3\partial_{z}a/{a}+\partial_{z}f_{\mathcal{R}}/{f_{\mathcal{R}}}$. The above equation has the form of $\mathcal{Q}^{\dag}\mathcal{Q}\Psi(z)=m^{2}\Psi(z)$, which ensures that the eigenvalues are nonnegative, i.e., $m^2 \ge 0$. Thus, there are no gravitational tachyon modes and the system is stable under tensor perturbations.
It should be pointed out that this is valid for any $f(\mathcal{R})$ with $f_{\mathcal{R}}>0$.

\subsection{Localization of massless graviton}

Now we analyze the localization of the graviton zero mode $\Psi_{0}(z)$, for which $m=0$ and the equation is reduced to  $\left(-\partial_{z}
 + \mathcal{A}/2\right)\Psi_{0}(z)=0$. The solution of the zero mode is
\be
 \Psi_{0}(z) \propto a^{3/2}f_{\mathcal{R}}^{1/2}. \label{zeromode}
\ee
Clearly, for the first constant curvature solution (\ref{AdSsolution}), the zero mode cannot be localized on the brane. For the second constant curvature solution (\ref{dSsolution}), though the zero mode can be localized on the brane, it can be shown that the energy density diverges at the boundaries of extra dimension. Now we focus on the nonconstant curvature solutions.
For the $f(\mathcal{R})=\mathcal{R}+\alpha \mathcal{R}^2$ brane model with $u=c_1 a^n$, the warp factors are given by $a(y)=\text{sech}^{\frac{2}{3(n-1)}}(ky)$ and $u(y)=c_1\text{sech}^{\frac{2n}{3(n-1)}}(ky)$ with $n>1$ or $n<-{2}/{3}$. Recalling Eq.~(\ref{definition equation of q}), the zero mode (\ref{zeromode}) is given by
\be
 \Psi_{0}(z(y)) \propto \text{sech}^{\frac{n}{n-1}}(ky).
\ee
The normalization condition for the zero mode is
\be
 \int^{+\infty}_{-\infty} \Psi_0^2 dz =\int^{+\infty}_{-\infty} \Psi_0^2 a^{-1}dy
  \propto \int^{+\infty}_{-\infty}  \text{sech}^{\frac{2(3n-1)}{3(n-1)}}(ky) dy < \infty,
\ee
which can be guaranteed for both the solutions with $n>1$ and $n<-{2}/{3}$. So the zero mode can always be localized on the brane and the four-dimensional gravity can be recovered on the brane.

\subsection{{Corrections to Newtonian potential of massive KK modes}}
\label{subsecMassiveKKmodes}
For massive KK modes, we need to analyze the properties of the effective potential $\mathcal{W}(z)$ given in Eq.~(\ref{Wz}) for different values of $n$. For the $f(\mathcal{R})=\mathcal{R}+\alpha \mathcal{R}^2$ brane model with $u=c_1 a^n$, the effective potential is reduced to
\be
 \mathcal{W}(z)=  \frac{3 n}{2}\frac{ \partial_z^2 a}{a}+ \frac{3 n(3 n-2)}{4}  \left(\frac{\partial_z a}{a}\right)^2,
 \label{Wz}
\ee
or
\ba
 \mathcal{W}(z(y)) &=& \frac{3n}{4}  \left( 2 a \partial_y^2 a+3 n (\partial_y a)^2\right) \nonumber \\
    &=& \frac{ n k^2}{3 (n-1)^2}
         \text{sech}^{\frac{4}{3 (n-1)}}(k y)
         \left[3 n+2 - (6 n-1) \text{sech}^2(k y)\right].
 \label{Wzy}
\ea
From the above equation, we can see that
\ba
 \mathcal{W}(0) &=& -\frac{ nk^2}{n-1},\label{W0} \\
 \mathcal{W}(|y|\rightarrow \infty)
   &\rightarrow& \frac{ n (3 n+2)k^2 }{3 (n-1)^2} e^{-\frac{4 k |y|}{3 (n-1)}}.
 \label{Wzinfty}
\ea

\begin{figure*}[htb]
\begin{center}
\includegraphics[width=7.5cm,height=5cm]{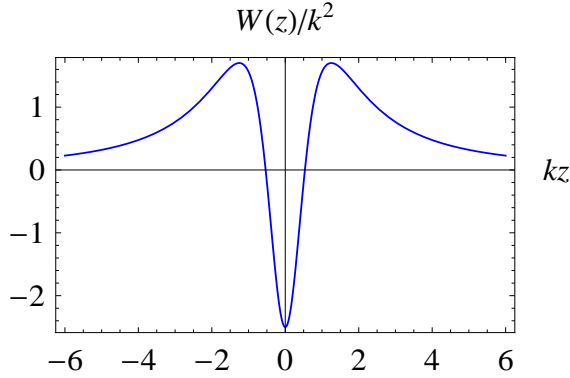}
\end{center}
\caption{The shape of the effective potential $\mathcal{W}(z)/k^2$ for the graviton KK modes with respect to $kz$ for $n= {5}/{3}$.} \label{figWz}
\end{figure*}

For $n>1$, the effective potential has a trapping well around the brane and trends to vanish from above at infinity,
 which shows that the effective potential has a trapping well around the brane and a potential barrier at each side of the brane.
The asymptotic behavior of the effective potential implies that only the zero mode is a bound state and any massive mode cannot be localized on the brane.
For $n={5}/{3}$, the explicit expression of the effective potential in the conformally flat coordinate $z$ can be obtained:
\ba
 \mathcal{W}(z) &=& \frac{5 k^2 ( 7 k^2 z^2-2)}{4 \left[(kz)^2+1\right]^2}.  \label{Wz2}
\ea
We show the plot of $\mathcal{W}(z)$ in figure~\ref{figWz}.

 As can be seen
 from Eq.~(\ref{Wz2}), the effective potential
 allows a series of continuous massive KK modes $\Psi_{m}$.
 They are not localized on the brane. As claimed in Ref.~\cite{Csaki2000a},
 if the effective potential $\mathcal{W}(z)\sim {\beta(\beta+1)}/{z^2}$ as $|z|\rightarrow\infty$,
 then the Newtonian potential is corrected by $\Delta U(r)\sim\ 1/r^{2\beta}$.
 For a general parameter $n$, the parameter $\beta$ is determined by $\beta(\beta+1)={n(3n+2)}/{3(n-1)^2}$, which gives
\ba
 \beta=\frac{\sqrt{3(15n^2 + 2n +3)}}{6(n-1)}-\frac{1}{2} ~~\Big(>\frac{\sqrt{5}-1}{2} \Big).  \label{beta}
\ea
 So in the case of $n= {5}/{3}$, the Newtonian potential is corrected by $\Delta U(r)\sim1/r^5$.
 Note that our result is different from the one obtained for the metric $f(R)=R+\alpha R^2$ brane model in Ref. \cite{Liu2011a}, which gives $\Delta U(r)\sim1/r^3$.

 What is more interesting is the case of $n<-{2}/{3}$. According to Eqs. (\ref{Wzy}) and (\ref{W0}), we can see that $\mathcal{W}(z(y))$ diverges as $|y|\rightarrow
 \infty$ and so $\mathcal{W}(z)$ diverges at the boundaries. Thus, all of the massive KK modes are bound and discrete states. However,
 it can be seen that the conformally flat coordinate
 $z=\int_{0}^y \text{sech}^{{2}/{3(1-n)}}(k \bar{y}) d\bar{y}$  ranges from $-z_0/2$ to $z_{0}/2$, where $z_{0}$ is a finite parameter  determined by $k$ and $n$. Hence $\mathcal{W}(z)$ acts as an infinitely deep potential well, which supports the solutions of highly excited states with even function $\Psi_{\text{even}}\simeq \sqrt{2}\text{cos}(mz)/{\sqrt{z_0}}$ and odd function $\Psi_{\text{odd}}\simeq \sqrt{2}\text{sin}(mz)/{\sqrt{z_0}}$. Indeed, the corrections to Newtonian potential of the KK modes can be roughly calculated by the infinitely deep square potential well model. The mass spectrum determined by the even KK modes $\Psi_{\text{even}}(z)= \sqrt{2}\text{cos}(m_{N} z)/{\sqrt{z_0}}$ is given by
\ba
 m_N = (2N+1)\pi/z_0 ,~~~  (N=1,2,3\cdots).
\ea
 With this mass spectrum, if we consider two test particles separated by a distance $r$ on the brane, then the Newtonian potential is~\cite{Csaki2000a,Guo:2010az}
\be
 U(r)=-\frac{M_1 M_2}{M_{\text{pl}}^2}\frac{1}{r} - \frac{M_1 M_2}{M_{*}^{3}}
 \sum_{N=1}^{\infty}\frac{e^{-m_N r}}{r}|\Psi_{m_N}(0)|^2,
 \label{correction 1}
\ee
where $M_{\text{pl}}$ is the Planck scale and $M_1$, $M_2$ are the masses of the two test particles. From the action (\ref{action of Pal. f(R)}), to recover the four-dimensional gravity, we have
\be
\frac{M_{*}^{3}}{2} \int d^5 x \sqrt{-g}f(\mathcal{R}(g,\Gamma))
\supset \frac{M_{\text{pl}}^2}{2}\int d^4 x \sqrt{-g^{(4)}(x^{\mu})} R^{(4)}(x^{\mu})
\ee
 with $g^{(4)}(x^{\mu})$ the determinant of the four-dimensional metric and $R^{(4)}(x^{\mu})$ the Ricci scalar in four dimensions. Therefore, the relation between the effective Planck scale $M_{\text{pl}}$ and the fundamental scale $M_{*}$ is given by
\be
 M_{\text{pl}}^2=M_{*}^{3} \int_{-\infty}^{+\infty} d y a^{2}(y)f_{\mathcal{R}}
 \equiv \frac{M_{*}^{3}}{k}\sigma_{1}(n),
 \label{de sigma1}
\ee
where $\sigma_{1}(n) \simeq \frac{3(n-1)(6n-1)}{2(3n-1)(3n+2)}$. For convenience, we define another function $\sigma_{2}(n)$ by
\be
z_0= 2\int_{0}^{+\infty} a^{-1}(y)d y\equiv \frac{2}{k}\sigma_{2}(n).
\label{de sigma2}
\ee
It is turned out that $\sigma_{2}(n)\simeq \frac{3}{2}(1-n)$.
In terms of $\sigma_{1}(n)$, $\sigma_{2}(n)$, and the relation (\ref{de sigma1}), the Newtonian potential (\ref{correction 1}) can be expressed as
\ba
U(r)&=&-\frac{M_1 M_2}{M_{\text{pl}}^2}\frac{1}{r}\left[1 + \frac{\sigma_{1}(n)}{\sigma_{2}(n)}
 \sum_{N=1}^{\infty}e^{-m_N r}\right]\nn\\
 &=&-\frac{M_1 M_2}{M_{\text{pl}}^2}\frac{1}{r}\left[1 + \frac{\sigma_{1}(n)}{\sigma_{2}(n)}
 \frac{e^{-{3\pi r}/{z_0}}}{\left(1-e^{-{2\pi r}/{z_0}}\right)}\right]\nn\\
 &\simeq&-\frac{M_1 M_2}{M_{\text{pl}}^2}\frac{1}{r}\left[1 + \frac{(6n-1)}{(1-3n)(3n+2)}
 \frac{e^{-{3\pi r}/{z_0}}}{\left(1-e^{-{2\pi r}/{z_0}}\right)}\right].
 \label{correction 2}
\ea
Clearly, for $r\gg z_0$, the summation term tends to $e^{-{3\pi r}/{z_0}}$ and thus can be ignored. For the case of $r\ll z_0$, we can expand the correction term in terms of $r/z_0$, and effective Newtonian potential is
\be
 U(r) \simeq -\frac{M_1 M_2}{M_{\text{pl}}^2}\frac{1}{r}\left[
 1+\frac{\sigma_{1}(n)}{{2\pi}\sigma_{2}(n)}  \frac{{z_0}}{r}\right],
\ee
which can also be expressed in a more elegant formalism by recalling  Eqs.~(\ref{de sigma1}) and (\ref{de sigma2}):
\be
U(r) \simeq -\frac{M_1 M_2}{M_{\text{pl}}^2}\frac{1}{r}
 -\frac{M_1 M_2}{M_{*}^3}\frac{1}{\pi r^2}.
\ee
It shows that the Newtonian potential is corrected by a $1/r^2$ term when $r\ll z_0$, and the correction term dominates, namely $U(r)\propto 1/r^2$. For $r\gg z_0$, the correction can be ignored, and we have $U(r)\propto 1/r$. This result is similar to the ADD model~\cite{Arkani-Hamed1998a}. In contrast to the ADD model, the physical extra dimension in our model is infinitely large.  The length parameter $z_0$ can be regarded as a length scale beyond which the four-dimensional gravity can be recovered.
We get this result by using the infinitely deep square potential model. One can also consider the potential model
$\bar{\mathcal{W}}(z)=c\text{ tan}^{2}({\pi}z/{z_0})$, and this would lead to a similar result.

\section{Discussions and conclusions}
\label{secConclusion}

 In this work we investigated the thick brane configuration generated by a background scalar field in Palatini $f(\mathcal{R})$ gravity. In this gravity, higher derivatives of the matter fields are involved in the field equations, whereas it is the metric that contains higher derivatives in metric $f(R)$ gravity. This leads to the difference of strategies to solve differential equations. For the case of constant curvature, the solutions are the same as in metric $f(R)$ gravity. The $dS_{5}$ solution supports an energy density diverges at infinity, which implies that it is not a viable brane solution. For nonconstant curvature, it is convenient to introduce an auxiliary metric to reduce the order of the differential equations, which is a key step in Palatini theories~\cite{Banados2010, Liu2012a, Fu2014a}. By assuming the relation between the spacetime metric and the auxiliary metric, we obtained the thick brane solutions of this system. The scalar field solution is a kink, which connects two vacua of the scalar potential, and it describes a domain wall brane. Besides, the thickness of the brane is determined by the coefficient of the $\mathcal{R}^2$ term.

 Furthermore, we analyzed the gravitational fluctuations of the brane system. For the TT tensor perturbations, a Schr\"odinger-like equation was obtained. It was shown that the Palatini $f(\mathcal{R})$-brane system with any function $f(\mathcal{R})$ with $f_\mathcal{R}>0$ is stable under tensor perturbations. For the $AdS_{5}$ solution, the graviton zero mode cannot be localized on the brane. The $dS_5$ solution suffers from the pathology that the energy density diverges at the boundaries of the extra dimension. For the nonconstant curvature solutions, the asymptotic behavior of the effective potential implies that the graviton zero mode can always be localized on the brane. The behaviors of massive KK modes are determined by the parameter $n$ in the brane solutions. For the solution with $n>1$, the effective potential is volcano one, so the massive KK modes are not bound states and they are suppressed on the brane due to the potential barrier near the brane. The more interesting case is the solution with $n<-2/3$, for which the effective potential is an infinitely deep potential well and the massive KK modes are all bound states. In this case, we get the correction to Newtonian potential, $\Delta U(r)\sim 1/r^2$, and the length scale beyond which the four-dimensional gravity can be recovered.
 This result is similar to the ADD model.

 At last, it should be pointed out that some of the massive KK modes in the case of $n>1$ may be quasilocalized on the brane due to the potential barrier at each side of the brane~\cite{Fu2014a, Liu2009}. It would be interesting to explore the properties of these quasilocalized states. Unfortunately, no such states were found in this model. We expect that these quasilocalized gravitons appear in some more general Palatini theories.

\section{Acknowledgement}
 This work was supported by the National Natural Science Foundation of China (Grants No. 11075065 and No. 11375075) and the Fundamental Research Funds for the Central Universities (Grant No. lzujbky-2015-jl01).
%

\end{document}